# Single Layer In-O Atomic Sheets as Phonon and Electron and Barriers in ZnO-In$_2$O$_3$ Natural Superlattices: Implications for Thermoelectricity

Xin Liang[1,3], David C. Bell[1,2] and David R. Clarke[1]

[1]School of Engineering and Applied Sciences, Harvard University, Cambridge, MA 02138, USA

[2]Center for Nanoscale Systems, Harvard University, Cambridge, MA 02138, USA

[3]Now at: School of Materials Science and Engineering, Changzhou University, Jiangsu 213164, China.

**ABSTRACT**

The phases in the ZnO half of the ZnO-In$_2$O$_3$ binary system are natural superlattices consisting of periodic stacking of single InO$_2$ sheets separated by indium doped ZnO blocks with a spacing that depends on the composition according to the relationship In$_2$O$_3$(ZnO)$_k$. Characterization by combined, atomic resolution, aberration-corrected scanning transmission electron microscopy (STEM) and electron energy loss spectroscopy (EELS) analysis, indicates that the atomic structure of each InO$_2$ layer consists of a single continuous sheet of octahedrally-coordinated InO$_2$. The sheets also are crystallographic inversion boundaries. Analysis of the electrical conductivity, thermal conductivity and Seebeck coefficient data at 800 °C indicates that the InO$_2$ sheets not only decrease thermal conductivity by phonon scattering but also create potential barriers to electron transport. The origin of the potential barriers, the role of piezoelectric effects and their dependence on superlattice spacing is discussed qualitatively. It is also argued that the crystallographically-aligned InO$_2$ sheets within individual grains are also transport barriers in randomly oriented polycrystalline materials.

I.    INTRODUCTION

Nanostructure engineering has been exploited in recent years to create thermoelectric materials with high figures of merit. This approach, which has included decreasing the grains size of polycrystalline materials, has arguably been most effective in semiconductor superlattices structures [1-4] and can be attributed to the high density of coherent interfaces that are effective in scattering phonons, dramatically decreasing thermal conductivity, while also modifying the density of states at the Fermi level to produce an increase in the Seebeck coefficient [5]. These superlattices as well as the crystallographically incoherent superlattices, such as W/alumina, that exhibit very low thermal conductivities [6-8] are all produced layer-by-layer, for instance, by molecular beam epitaxy (MBE) [6, 9]. Furthermore, these synthetic superlattice structures are intrinsically morphologically unstable at high temperatures due to their high interfacial energy [7, 9] making them unsuitable for prolonged, high temperature applications.

A number of oxide semiconductors form superlattices naturally. These include $In_2O_3(ZnO)_k$ [10-17], $CaO(CaMnO_3)_k$ [18, 19], $Na_xCo_2O_4$ [20, 21] and $Ga_2O_3(ZnO)_k$ [22-24], where the integer $k$ denotes the order of the superlattice structure. In the ZnO-based superlattices the integer $k$ describes the number of ZnO planes between the indium oxide layers [13, 17, 25] measured along the c-axis of the ZnO. The superlattices in these materials form from about 10 mole percent (m/o) to 40 m/o $InO_{1.5}$. As they self-assemble on annealing at high temperatures, they can be made in bulk by sintering powders of the appropriate composition with their spacing determined directly by their overall composition. (For compositions less than about 10 m/o $InO_{1.5}$, individual indium ions dissolve into the ZnO blocks substituting for the $Zn^{2+}$ ions. Such solid solution appears to entropically favored and is responsible for the observed increase in n-type electrical conductivity of ZnO when doped with $In^{3+}$ ions. For compositions greater than 40 m/o $InO_{1.5}$, the materials evolve into a two-phase system consisting of a mixture of $In_2O_3$ and a natural superlattice. The properties of these two-phase materials will not be discussed here since the $In_2O_3$ phase grains can act as a parallel conducting path, both electrically and thermally). As the compounds are thermodynamically stable, the superlattices are, unlike grain or precipitate microstructures, resistant to coarsening at high temperatures [26] and remain morphologically stable [17].

In this publication we provide detailed atomic structure of the superlattice interface, present data derived from the Seebeck coefficient on the electron barrier heights of the interfaces as a function of composition across the ZnO-In$_2$O$_3$ phase diagram, and relate them to the thermoelectric figure of merit previously reported for different superlattice compositions. The property data presented is for 800 °C in air as we are interested in high temperature thermoelectrics. Also, although these materials do not exhibit remarkable values of the figure of merit but it is anticipated that the systematic variation in properties with superlattice spacing presented in this work may nevertheless provide new insights.

## II. EXPERIMENTAL DETAILS

The ZnO-In$_2$O$_3$ powders were synthesized using combined wet chemistry and gel combustion methods and then sintered into solid pellets using a current-assisted densification processing technique [17, 26]. The materials made this way are bulk, randomly oriented, polycrystalline materials with a grain size of several microns [17, 26, 27]. Post-annealing treatments at 1250 °C for seven days were then given to the materials to stabilize the superlattice spacings. At this temperature, the phase diagram and the superlattices are well documented. Thermal conductivity measurements were made using the laser flash method up to 1000 °C. Measurement of the electrical conductivities and Seebeck coefficient were made using a ULVAC RIKO ZEM 3 M10 unit. Complementary measurements from room temperature up to 800 °C have previously been reported [17, 26, 27].

## III. ATOMIC STRUCTURE OF THE SUPERLATTICE INTERFACE

Since their discovery, there has been some ambiguity as to the detailed atomic structure of the superlattice interfaces in the $In_2O_3(ZnO)_k$ system. Early on, the atomic structure was described as consisting of alternating, periodic stacking of blocks of $In_2O_3$ and $k$-unit cells of ZnO [10]. More recently the atomic structure of these superlattice compounds has been revised and is now described as consisting of a periodic arrangement of single, octahedrally coordinated $InO_2$ layer separating $(k + 1)$ unit cells of ZnO stacked along the ZnO [0001] c-axis. This has been substantiated by crystal structure refinement [28] and by some high-resolution electron microscopy observations [14, 29, 30] as well as 3D atom probe tomography [26]. In this study, the atomic structure is unambiguously resolved as a single $InO_2$ layer using the atomic resolution aberration-corrected STEM angular dark field (ADF) imaging mode in a Hitachi HD-2700 microscope operating at 200 keV. An example is shown in Figure 1 (a) where the $InO_2$ sheets arranged in between blocks of ZnO are seen edge-on in a 10 m/o $InO_{1.5}$ doped ZnO post-annealed at 1250 °C for 7 days. In the ADF imaging mode, indium, having higher atomic mass, $Z$, than zinc, appears significantly brighter. Oxygen ions are not seen due to their much smaller atomic mass. (In ADF imaging the contrast varies as $Z$-squared). The observed atomic structure by aberration corrected STEM imaging confirms the atomic models determined by X-ray diffraction [25, 31, 32] and recent microscope based spectroscopy work by Schmid *et al.* [29, 30] which suggests that the cation sites in $In_2O_3(ZnO)_k$ superlattice interface are solely occupied by indium ions that sit on octahedral sites coordinated to oxygen ions. The localization of the indium ions is confirmed by EELS spectra recorded from adjacent atomic positions, one on the bright column of atoms and the other on the adjacent weak column of atoms, shown in Figure 1 (b) and (c), respectively. The spectrum recorded from a bright atom clearly reveals that it is In and O by the presence of the characteristic $In-M_{4,5}$, $In-M_3$ and O-K edges. The other spectrum recorded from an adjacent atomic position located off the $InO_2$ sheet has no discernable edges from Indium and only the $Zn-L_{2,3}$ and $Zn-L_1$ edges from the Zn in addition to the O-K edge. These measurements were confirmed by EELS spectra from successive atom positions across the InO$_2$ sheet, as shown in Figure 2.

Crystallographically, the InO$_2$ sheets comprising the superlattice interface are coherent, inversion domain boundaries (IDB) with the polar direction of the ZnO reversing across the boundary [14, 33]. This is seen in the atomic model of Figure 3 for the k=6 composition constructed using a VESTA program [34] with the input of atomic coordinate information from a ICSD incorporated ICDD database [35]. Although the atomic arrangement for the In$_2$O$_3$(ZnO)$_6$ compound is shown in this figure for illustration, all the superlattices, irrespective of their $k$ value, consist of $(k+1)$ ZnO layers separating individual single-layer InO$_2$ sheets. Half-way between the InO$_2$ sheets and parallel to them is a mirror domain boundary (MDB) in the ZnO where there is a single sheet of zinc atoms with trigonal rather than tetrahedral coordination with its neighboring oxygen ions. Sometimes referred to as a plane of bipyramidal sites since this describes the local atomic coordination, there is again a reversal of the polar direction in the ZnO across this mirror plane. According to the structure, this corresponds to a "tail-to-tail" change in polarity in the ZnO. The atomic structure of the superlattice in Figure 3 is confirmed by the STEM ADF image with image matching. This confirms that there is a reversal in polarity in the ZnO at the InO$_2$ layer and also one at the mid-plane. Although the ideal structure consists of a single periodicity of the InO$_2$ sheets throughout the bulk materials, our observations indicate that there is always some variability in the spacing from region to region, even within individual grains, suggesting that thermodynamic equilibrium has not completely attained. Within the majority of grains all the superlattices are parallel to one another and in some grains there appear to be sub-grains within which the superlattices are parallel to one another. In addition to this structural description of the superlattice, individual indium ions can also dissolve into the ZnO blocks substituting for the Zn$^{2+}$ ions and enhancing the electrical conductivity.

## IV. PHONON SCATTERING

As shown in Figure 4, the thermal conductivity at 800 °C is dramatically reduced once sufficient $In_2O_3$ is alloyed into pure ZnO to form superlattices [17, 26, 27]. At higher concentrations, above about 10 m/o $InO_{1.5}$, the thermal conductivity is almost independent of $In_2O_3$ concentration, having a value of ~ 2 W/mK, until $In_2O_3$ forms as a second phase at compositions above 40 m/o $InO_{1.5}$. Previous analysis of the thermal conductivity data as a function of superlattice spacing indicates that each $InO_2$ superlattice sheet has a thermal resistance, $R_k$, of $5.0 \pm 0.6 \times 10^{-10}$ m²K/W at temperatures above room temperature [26]. This value of thermal resistance is intermediate between those reported for epitaxial semiconductor interfaces ($\sim 10^{-10}$ m²K/W) and those for grain boundaries ($10 \sim 50 \times 10^{-10}$ m²K/W). The net thermal resistance perpendicular to the superlattice is then simply the sum of the thermal resistance of the In-doped ZnO blocks and the number of superlattice sheets in the unit length: $\frac{1}{\kappa} = \frac{1}{\kappa_{ZnO}} + \frac{R_k}{(k+1)d_{\{0002\}}}$. The origin of the thermal resistance is attributed to both the lattice distortions associated with the crystallographic inversion across the interface and also the significantly higher atomic mass of Indium (114.82 versus 65.38 for the mass of Zn). The grain size of the materials studied was several microns and so the thermal resistance of the grain boundaries can be neglected.

One explanation for the low and compositional independence of the thermal conductivity is that the superlattice causes zone folding, restricting the acoustic velocity. Direct evidence for zone folding by the superlattice structure comes from Raman and vibrational studies of materials across the ZnO-$In_2O_3$ phase diagram [36]. Over the compositional range at which the superlattices form, low energy Raman bands appear. Furthermore, their Raman shifts correspond to those expected based on zone folding of the acoustic modes along the c-axis of the ZnO, as shown in Figure 5 taken from reference 36. In the same superlattice compositional range, new vibrational modes also appear and there is no evidence for the Raman modes of either ZnO or $In_2O_3$, the terminal phases of the binary phase diagram.

## V. ELECTRICAL CONDUCTIVITY AND SEEBECK COEFFICIENT

Over the compositional range that superlattices are observed to form, the electrical conductivity is almost independent of temperature from room temperature to 800 °C, as shown in Figure 6. This is in marked contrast to the usual thermally activated electrical conductivity of semiconductors, exhibited, for instance in lightly doped ZnO doped with less than 0.03 a/o Al [37, 38] or, for instance, with other solid solution dopants. It is also not observed in our materials for doping below 10 m/o $InO_{1.5}$. The observed conductivity increases with the $InO_{1.5}$ concentration as might be expected but whether this is due to an increase in overall doping alone or due to increases in mobility would require high-temperature Hall measurements. It is also not observed in polycrystalline ZnO-based varistors, which have cobalt-solid solution doping, and large ions, such as Bi and Pr, segregated to their grain boundaries. Consequently, this is believed to be a characteristic of electrical transport in these polycrystalline superlattice materials. Temperature-independent electrical conductivity is also one of the signature characteristics of tunneling controlled conductivity although we are unable to demonstrate experimentally that this is the case in these materials. Measurements of the Seebeck coefficient at 800 °C over the same compositional range are also shown in Figure 4. The data indicates that there is an almost linear decrease from - 450 µV/K at 10 m/o $InO_{1.5}$ to – 180 µV/K at 40 m/o $InO_{1.5}$ for the $k = 3$ superlattice. These values compare with a value of – 350 µV/K for the solid solution ZnO with 3 m/o $InO_{1.5}$.

## VI. POTENTIAL BARRIER HEIGHT

The atomic structure of the superlattice interface and the reversal in polarity of the ZnO blocks on either side of the interface suggests that there may be a potential barrier at each $InO_2$ superlattice interface. Consequently, electrons with energies below the superlattice interface barrier height, $\varepsilon_b$ , will have a high probability of being scattered whereas those with excess energies will not be scattered, decreasing the electron conductivity within a grain. Such "energy filtering" can also increase the Seebeck coefficient, $S$ [5]. For a spatially homogeneous, degenerate $n$-type semiconductor, the Seebeck coefficient can be expressed in terms of the electron density of states at the Fermi level by the relationship [39]

$$S = \frac{1}{eT}\left[\frac{\int_0^\infty \tau(\varepsilon)g(\varepsilon)\varepsilon^2\left(-\frac{\partial f(\varepsilon)}{\partial \varepsilon}\right)d\varepsilon}{\int_0^\infty \tau(\varepsilon)g(\varepsilon)\varepsilon\left(-\frac{\partial f(\varepsilon)}{\partial \varepsilon}\right)d\varepsilon} - \mu\right] \quad (1)$$

where $e$ is the electronic charge, $T$ is the absolute temperature, $\varepsilon$ is the electron energy, $\tau(\varepsilon)$ is the relaxation time, $g(\varepsilon)$ is the density of states, $f(\varepsilon)$ is the Fermi distribution function and $\mu$ is the Fermi energy. When superlattice interfaces are present, the electrons with energies below the superlattice interface potential barrier height $\varepsilon_b$ are strongly scattered.

Because of the barriers, the equation for the Seebeck coefficient must be modified by replacing the lower limit of the integral with the potential barrier height, $\varepsilon_b$:

$$S = \frac{1}{eT}\left[\frac{\int_{\varepsilon_b}^\infty \tau(\varepsilon)g(\varepsilon)\varepsilon^2\left(-\frac{\partial f(\varepsilon)}{\partial \varepsilon}\right)d\varepsilon}{\int_{\varepsilon_b}^\infty \tau(\varepsilon)g(\varepsilon)\varepsilon\left(-\frac{\partial f(\varepsilon)}{\partial \varepsilon}\right)d\varepsilon} - \mu\right] \quad (2)$$

In the limit $\mu/k_BT \gg 1$, Kishimoto *et al.* [40] have shown that this equation can be approximated by:

$$S = \frac{k_B}{e}\left[\eta^* + (1 + \exp(\eta^*))\ln(1 + \exp(-\eta^*))\right] \quad (3)$$

where $\eta^* = (\varepsilon_b - \mu)/k_BT$, $k_B$ is the Boltzmann constant. This, in turn, suggests that the variation in Seebeck coefficient will be almost linear with the barrier height.

Using equation 3, the effective potential barrier heights ($\varepsilon_b - \mu$) for the superlattice interfaces across the ZnO-In$_2$O$_3$ binary phase field can be estimated from the data for the Seebeck coefficient and electron conductivity [17, 27]. This is shown, in Figure 7, for the data at 800 °C as a function of indium oxide concentration. There is a strong dependence on composition, and, equivalently, on the superlattice spacing $(k + 1)d_{\{0002\}}$. In Figure 7, the barrier height is superimposed on the same graph as the experimental data for the figure of merit, $ZT$:

$$Z^*T = \frac{S^{*2}\sigma^*}{\kappa^*} T \qquad (4)$$

where the superscript * indicates the polycrystalline, bulk measurement. Comparison of the cross-plots in the figure indicates an inverse correlation between the two. In part this is because the thermal conductivity (figure 4) is almost independent of composition and so the figure of merit is proportional to the $S^{*2}\sigma^*T$ product. Taken together with the increase in Seebeck coefficient and the fact that the greatest change in the individual parameters with composition the conductivity, this would suggest that $Z^*T$ is controlled by an interplay between barrier scattering of electrons and the barrier height determining the Seebeck coefficient. Accordingly, there should be an optimal effective barrier height that maximizes the power factor and therefore ZT leading to the possibility of tuning both the power factor and ZT through composition control.

Assuming that the barrier height is the same for each $InO_2$ sheet, then in compositions where their spacing is sufficiently large so that there are no interactions between the adjacent $InO_2$ sheets, the value of the barrier height can be extracted from the variation in the effective barrier height with indium concentration can be considered to be an increase with the superlattice spacing. Thus, extrapolating the linear dependence of the barrier height to pure ZnO, corresponding to the limit of $InO_2$ sheets infinitely far apart, the barrier height would have a value of 0.49 eV.

## VII. DISCUSSION

Electrical and thermal transport in polycrystalline materials in which there are superlattices in the individual grains is complicated and it is recognized that it cannot be fully understood on the basis of polycrystalline measurements alone. For instance, to what extent can electrons transport in the ZnO blocks alone, cross the superlattice interfaces and can they also transport in the $InO_2$ sheets themselves as a parallel path. Furthermore, to what extent do the grain boundaries affect the measurements. Strictly speaking answering these questions requires high-temperature studies of single crystal but they do not exist. Nevertheless, some progress towards addressing these questions can be made by examining the systematic variation in properties with the superlattice spacing and by making the assumption that as the grains sizes are orders of magnitude larger (microns) than the superlattice spacings (nm), the thermal resistance of the grain boundaries can be neglected.

We start with the observations that the effective barrier height and electrical properties depend on composition, suggesting that it is likely that the barriers to electron transport are related to the $InO_2$ sheets themselves. It could be argued that electrons could flow between the sheets and not cross any barriers. But as will be shown later, in a polycrystalline material, they must do so. For this reason, our focus is on electrical transport across the $InO_2$ sheets. It is proposed in this discussion that the barrier is associated with the band offset between $InO_2$ and ZnO although, as will also be discussed, this may be modified by the polarization effects associated with the ZnO polarization.

Information regarding the band-gap of $In_2O_3$ is limited. There appears not to be a general consensus but according to recent calculations the band-gap has a value of 2.94 eV [41]. In contrast, there are numerous experimental data and *ab-inito* calculations indicating that the band-gap of ZnO is close to 3.4 eV. Fortunately, ZnO is known to be an *n*-type semiconductor and our conductivity data confirms that indium doping of ZnO in the superlattice regime increases the *n*-type conductivity. However, nothing is known about the conductivity of the $InO_2$ sheets themselves. Unless there are unidentified aliovalent impurities present, it is very likely that each is an intrinsic, albeit two-dimensional, semiconductor. As the Fermi levels of the $InO_2$ sheets and the ZnO blocks must coincide, this implies that the conduction band energy of the $InO_2$ sheets is higher than that of the ZnO thereby providing an intrinsic barrier to electron transport, as shown in Figure 8. Assuming that the Fermi level is close to the conduction band in the ZnO blocks, this would give an estimate of the barrier height of 0.46 eV. This estimate is consistent with the effective barrier height data in Figure 7 when extrapolated to large values of $k$, namely $InO_2$ sheets spaced far apart. The existence of a very narrow barrier due to the $InO_2$ sheets also provides a ready, albeit only qualitative, explanation for the unusual temperature dependence of the electrical conductivity of the superlattice structures shown in Figure 6.

This simple band-offset model does not, by itself, however, provide an explanation for one of the surprising results namely that the effective barrier height varies with composition, corresponding directly to the spacing of the $InO_2$ sheets. As each sheet, irrespective of their spacing, is chemically and structurally the same and they are separated by blocks of ZnO it is highly unlikely that the size of the block can directly affect the effective barrier height. Furthermore, although it is possible that the Fermi energy in the blocks varies with their size, any such effect should already be included since the effective barrier height is the difference between the top of the barrier and the Fermi level. However, this band offset is expected to be independent of material composition so an additional mechanism that affects the barrier height must be taken into account. We propose that the barrier height is affected by polarization effects associated with the fact that the $InO_2$ sheets are also crystallographically inversion interfaces in the superlattices.

As with other wurtzite compounds, such as GaN and AlN [42], it is known that the direction of spontaneous polarization in ZnO is opposite to the direction of its c-axis. (By convention, the c-axis positive vector points from a Zn ion to an O ion and is perpendicular to the basal planes (0001) of the wurtzite structure.) Both experiments and *ab-initio* calculations indicate that the spontaneous polarization of ZnO has a value of about $-0.05$ C/m$^2$ [42, 43]. Thus, contrary to the common assumption that the polarization direction points from the cation to the anion, the direction of spontaneous polarization in ZnO is opposite. This direction is indicated in the Figure 3. In response to this fixed electrostatic charge, an internal electric field will be created that causes tilting of the conduction and valence bands as shown schematically (and exaggeratedly) in Figure 8 with a maximum electron energy mid-way between the InO$_2$ sheets at the location of the MDB and sloping towards them. Depending on the value of the Fermi level, electrons will accumulate on both sides of the InO$_2$ sheets. Although the thickness of the InO$_2$ sheets is only of angstrom dimensions, the symmetry of the bands dictates that there will be no net tunneling current through the barrier in the absence of any voltage bias. Also, because the polarization charge is a constant, a feature of the polarization electric field is that it depends on the ZnO block size, namely the smaller the spacing of the InO$_2$ sheets the larger is the internal electric field. Consequently, the charge accumulation and the effective barrier height are expected to decrease as the InO$_2$ sheet spacing decreases. When the InO$_2$ sheets are far apart then the polarization field is negligible and the dominant barrier is the band offset discussed at the beginning of this section. The reversal in polarity at the mirror plane in the center of each ZnO block together with the direction of spontaneous polarization requires that it is a plane having a net positive charge. Correspondingly, there is a peak in the potential that electrons must overcome to move across the ZnO block and it is hypothesized that the effective barrier height is the difference between the mid-block potential and the band offset at the InO$_2$ sheets. A detailed band diagram calculation is clearly needed for a more complete and quantitative comparison with our observations but we consider that these polarization effects will be key factors in any explanation.

While the superlattice spacings in our material system are too small to enable direct measurements of the barrier height, single inversion twin boundaries have been reported to occur in each grain in ZnO varistors when $Sb_2O_3$ is included in the compositional mix. (The grain sizes in these varistor materials are typically several microns making the measurements straightforward). Voltage contrast imaging in the SEM, as well as micro-contact measurements on either side of these twin boundaries, indicate that there is a potential barrier when a voltage is applied across the inversion twin boundaries, where the Sb ions are located [44]. Based on the chemical and structural similarity between the $In^{3+}$ and $Sb^{3+}$ ions we believe that these observations of a potential barrier in the $Sb^{3+}$ doped ZnO are consistent with our suggestion that a barrier exists at our natural superlattices to transport across them.

As mentioned earlier, the materials we have studied were polycrystalline with no detectable crystallographic texture and all the measurements were made on these bulk materials. The superlattice created by the aligned $InO_2$ sheets and the associated potential distribution within the grains superimposes an additional anisotropy on the existing electron mobility anisotropy of the wurtzite ZnO. (The latter for pure ZnO is reported to depend on temperature at temperatures below room temperature but negligible above [45].). For this reason the thermoelectric figure of merit is related to the polycrystalline values of the properties in equation 4 and raises the question of how they are related to the single crystal values of the natural superlattice. Some insight into this question can be gained by considering the polycrystalline averages of second-rank conductivity tensor properties. These were detailed by Mityushov *et al*. [46] and subsequently validated by finite element calculations by Yang *et al*. [47]. For a second rank tensor conductivity, **A**, such as electrical conductivity, thermal conductivity and Seebeck coefficient, the tensor components are given by:

$$A = \begin{pmatrix} a_x & 0 & 0 \\ 0 & a_x & 0 \\ 0 & 0 & a_x \end{pmatrix} \qquad (5)$$

where the conductivity anisotropy can be expressed in terms of the ratio, $r = a_z/a_x$, the ratio of conductivity perpendicular to parallel the InO$_2$ sheets in this material superlattice system. The third invariant of the conductivity is, in turn, $a_m = (a_x^2 a_z)^{1/3}$. Mityushov *et al*. show that the effective conductivity, $a_{bulk}^*$, of a bulk material containing grains each with a single orientation of aligned superlattices and absent any texture is:

$$a_{bulk}^*/a_m = \frac{r^{2/3}}{3} + \left[\frac{2}{3} - \frac{2}{9}\left(\frac{(r-1)^2}{r+2}\right)\right]r^{-1/3} \qquad (6)$$

One of the key features of this expression is that while the precise value of the conductivity of the bulk material depends on the actual conductivity anisotropy ratio, the single crystal conductivity anisotropy always affects the bulk conductivity of a polycrystalline material. In the context of this work, it means that the potential barrier associated with electron transport across the InO$_2$ sheets will always affect the measured electrical conductivity and the Seebeck coefficient of the polycrystalline bulk material under steady state conditions. The only exception is when the conductivity parallel to the sheets is many orders of magnitude larger than the conductivity perpendicular to the sheets. Consequently, in randomly oriented, polycrystalline bulk material, electrons must have to pass over barriers associated with the InO$_2$ sheets: while they may effectively short-circuit some of them but cannot avoid them entirely. The same conclusion applies to the thermal conductivity as it is also a second-order rank tensor. Thus, the overall thermoelectric properties of the polycrystalline material are affected by the presence of the InO$_2$ sheets which serve as effective barriers to electron and phonon transport.

In summary, our work indicates that the InO$_2$ sheets in the In$_2$O$_3$(ZnO)$_k$ superlattices act to not only scatter phonons and decrease thermal conductivity but also alter the Seebeck coefficient and electrical conductivity. While the energy filtering is beneficial in increasing the Seebeck effect, in this particular system the compositions with the lowest barrier heights exhibit the largest thermoelectric figure of merit presumably because of the much larger electrical conductivity. Consequently, it is not possible in this system to simultaneously increase both electrical conductivity and Seebeck coefficients.

However, not explored in this work is the possibility of modifying the electrical properties of the $InO_2$ sheets by isovalent substitution, for instance, using $Fe^{3+}$ or $Ga^{3+}$.

In closing, as a wide variety of natural superlattice oxides are known to exist and their compositions can be tailored by ionic substitutions according to established crystal chemical rules, we believe that this class of materials offers the opportunity for a new direction in seeking high-temperature thermoelectrics. For instance, since the original version of this paper was written, one of the authors has shown that both the Kapitza resistance and the electron potential barrier of the polycrystalline $In_2O_3(ZnO)_9$ superlattice can be greatly increased by doping with $Al^{3+}$ [48]. Small concentrations of $Al^{3+}$ are known to substitute for $In^{3+}$ and, based on their chemical bonding characteristics, are expected to enter into the $InO_2$ sheets. What is especially intriguing about these recent findings is that the ionic substitution not only affects the potential barrier but also the thermal conductivity. The fact that similar natural superlattices have been reported to occur in the $ZnO-In_2O_3-Fe_2O_3$ and $ZnO-In_2O_3-Ga_2O_3$ systems [25, 49-51] as well suggests that substitutional studies in these systems may also be of interest. They may also clarify the role of piezoelectric contributions, if any, to the potential barrier height. Furthermore, the flexibility afforded by ionic substitutional chemistry in other natural superlattices, such as in the $CaO(CaMnO_3)_k$, broadens the potential for natural superlattices as stable, high-temperature thermoelectrics.

**ACKNOWLEDGEMENTS**

The authors wish to thank Takahiro Sato and Yoshihiro Ohtsu of the Hitachi High-Technologies Corporation, Electron Microscope Application Development Department, for their assistance with acquiring the high resolution STEM data. We are also grateful to Dr. Samuel Margueron for allowing us to use the data in Figure 5.

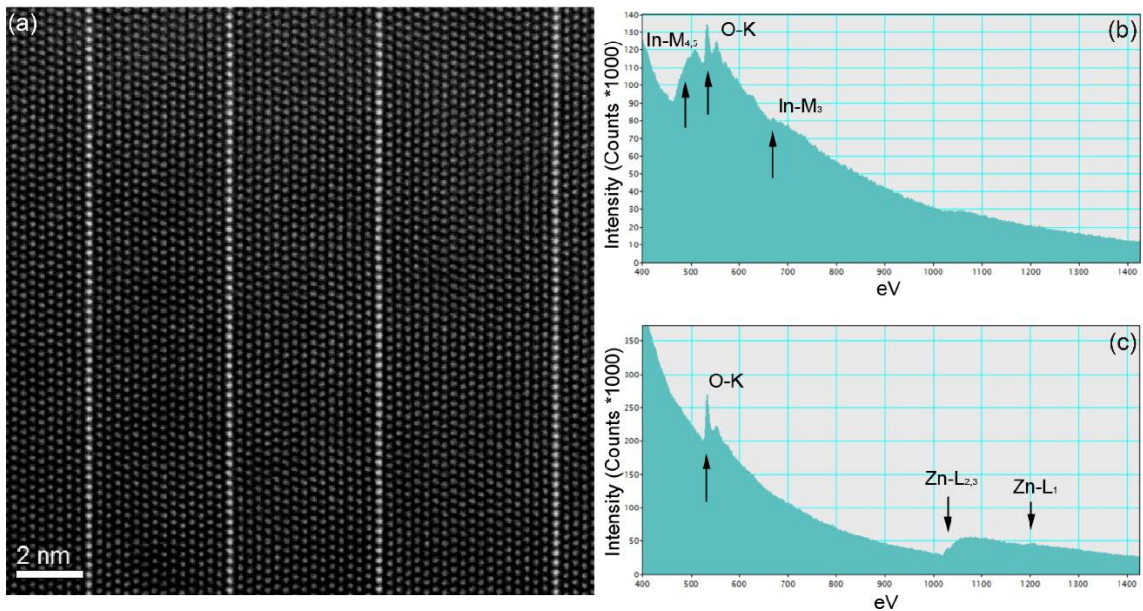

Figure 1: (a) Aberration corrected STEM ADF image of $In_2O_3(ZnO)_k$ natural superlattice structure showing the $InO_2$ sheets end-on and perpendicular to the c-axis of the ZnO. (b) EELS spectra from one of the bright columns of atoms and (c) an adjacent atom column together showing that the Indium ions are localized in the bright sheets in (a). In common with all the other materials in this work, this sample was annealed at 1250 °C for 7 days. This particular superlattice image is chosen as it indicates that there can be local variations in the superlattice spacing.

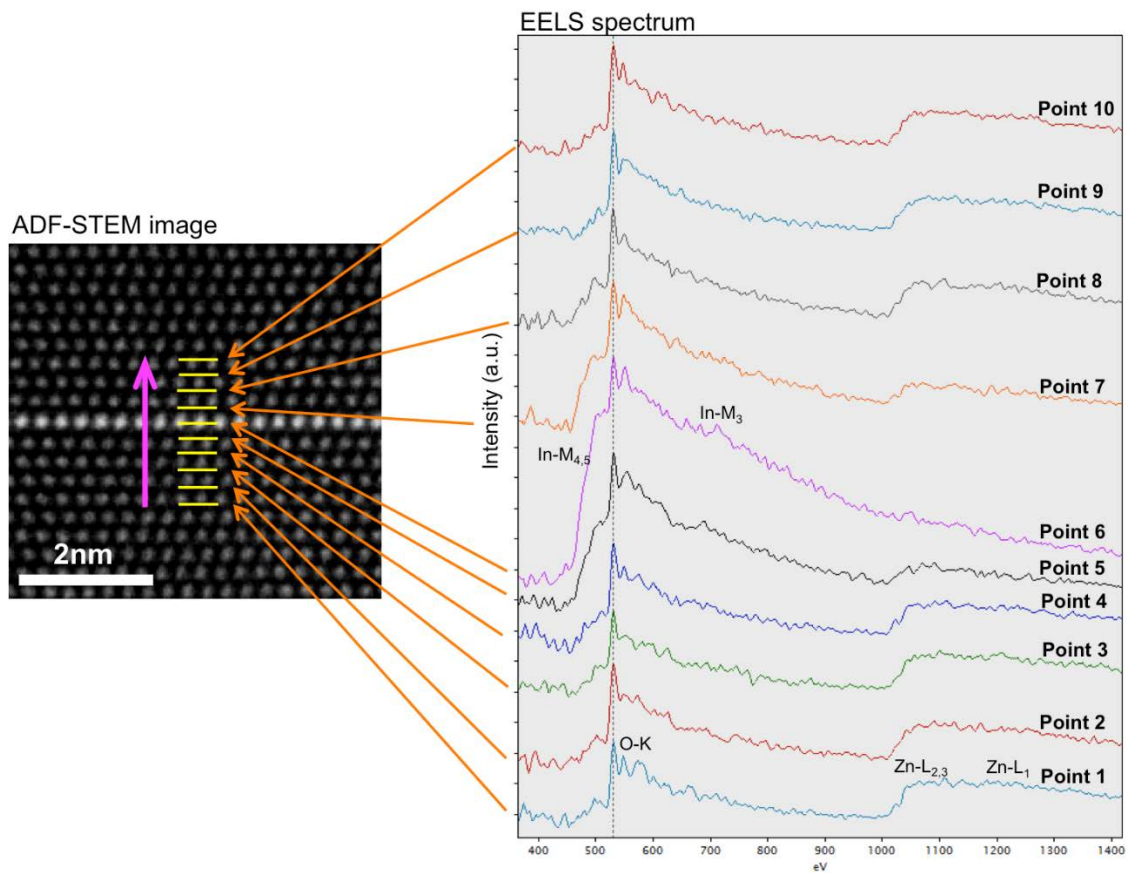

Figure 2. Atomic resolution electron energy loss spectra recorded at successive atomic positions across the one of the bright rows of atoms in Figure 1 (a). These spectra provide further confirmation that the Indium ions are indeed localized to the bright sheets imaged edge on as suggested by other TEM images and by the X-ray diffraction in literature.

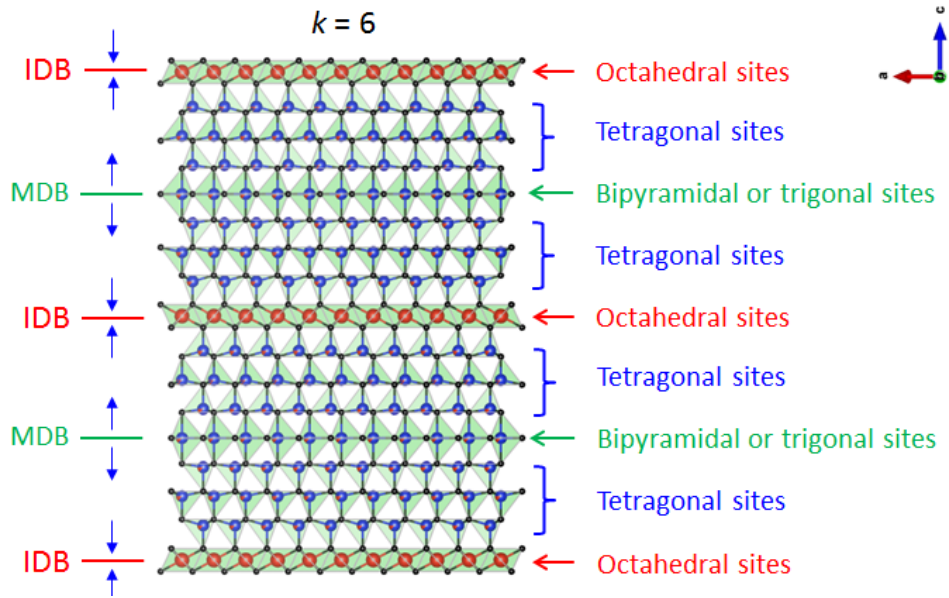

Figure 2: Atomic model of In$_2$O$_3$(ZnO)$_6$ superlattice compound viewed along the *b*-axis. Indium, zinc and oxygen ions are represented by red, blue and black balls respectively. The In-O and Zn-O bonding are visualized by the superimposed polyhedra. On the right side of the model, cation occupation sites are indicated; on left side, IDBs and MDBs are indicated and the direction of the positive c-axis vector in each ZnO block is represented by the blue arrow.

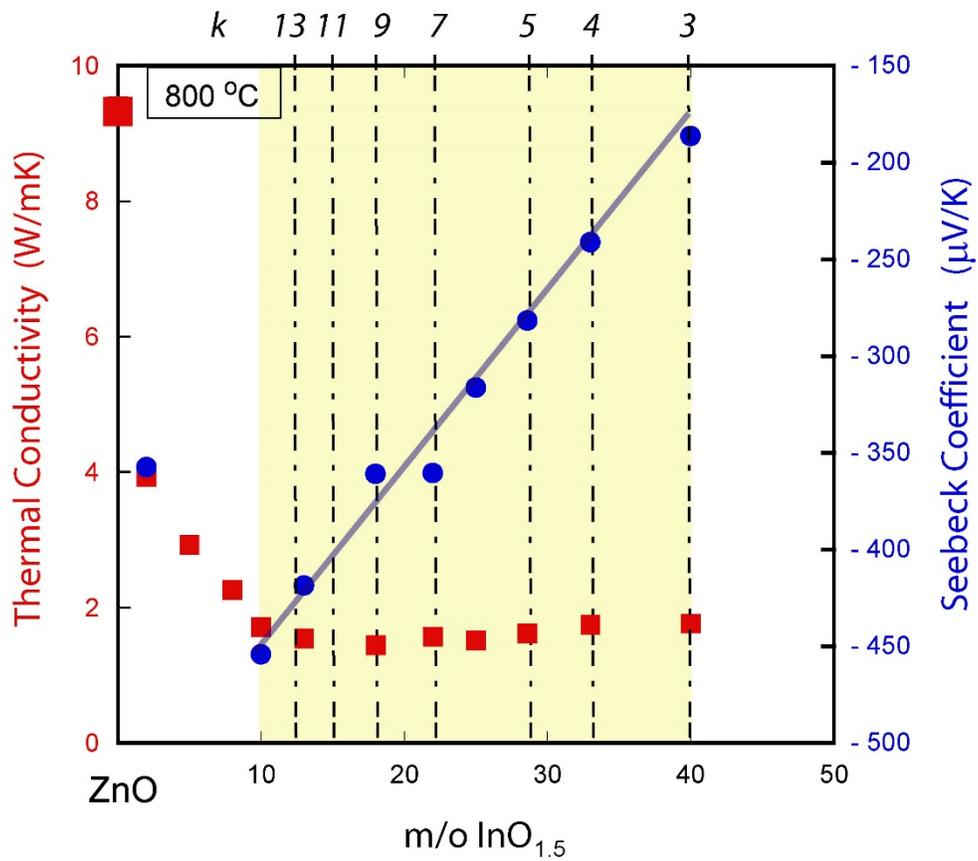

Figure 4. Variation of thermal conductivity and Seebeck coefficient with indium oxide concentration, all at 800 °C, over the compositional range where natural $In_2O_3(ZnO)_k$ superlattices form (shaded). The $k$ values corresponding to the compositions are labeled along top axis. Line though the Seebeck coefficient data is a guide to the eye.

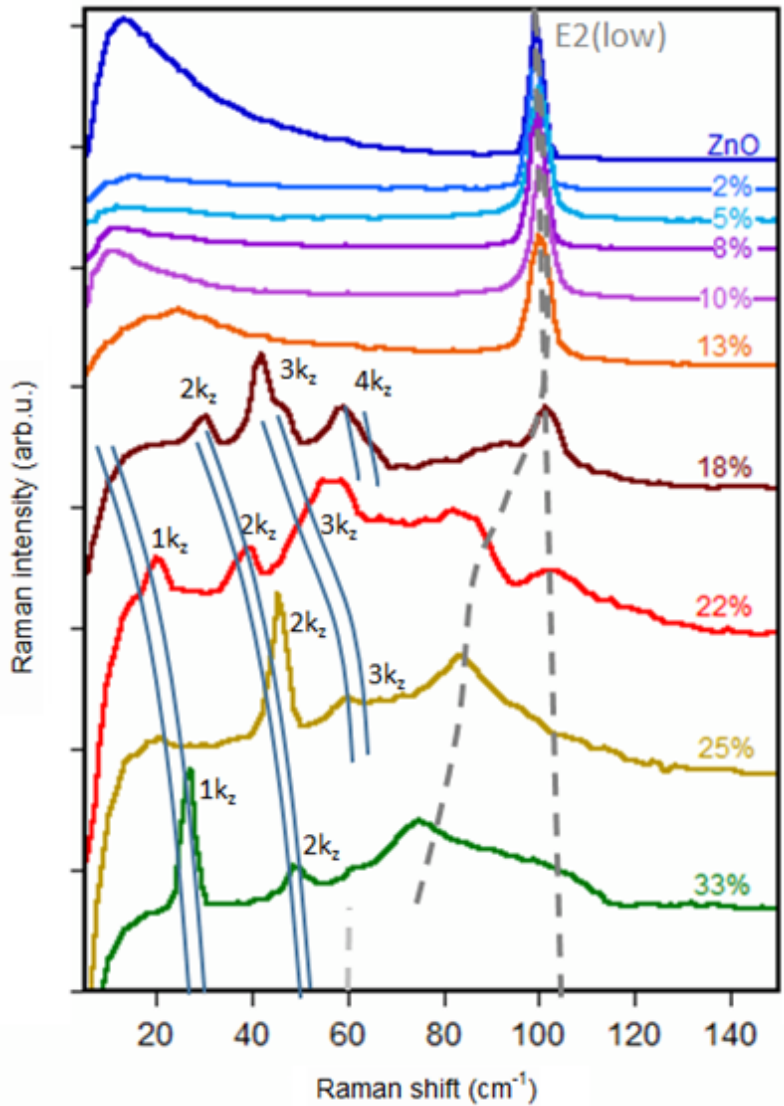

Figure 5. Low frequency Raman spectra for the compositions indicated. The Raman modes associated with zone folding of the acoustic branch of the superlattices along the c-axis are labeled and compared with calculated folding (the continuous curves $nk_z$). Reproduced from Margueron, *et al*.

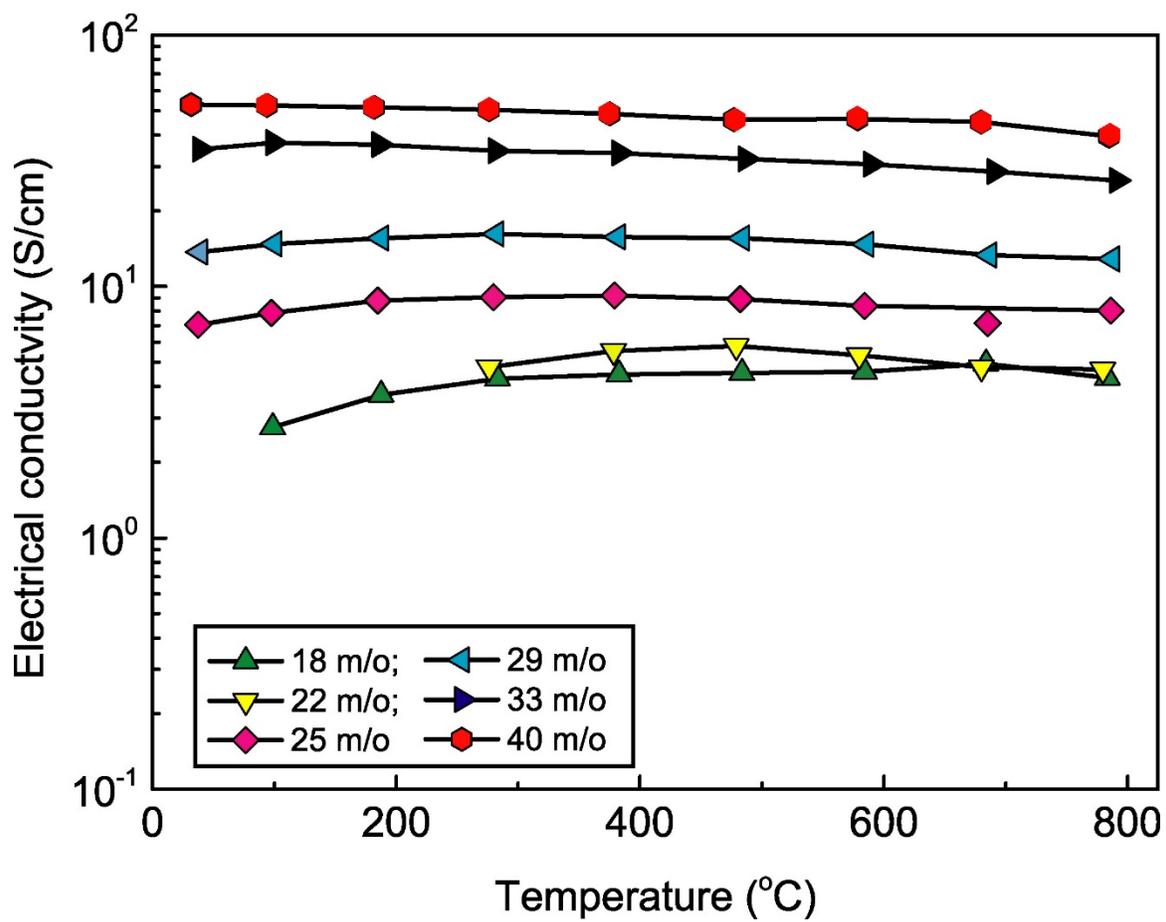

Figure 6. The electrical conductivity of each of the natural superlattice structures is almost independent of temperature above room temperature.

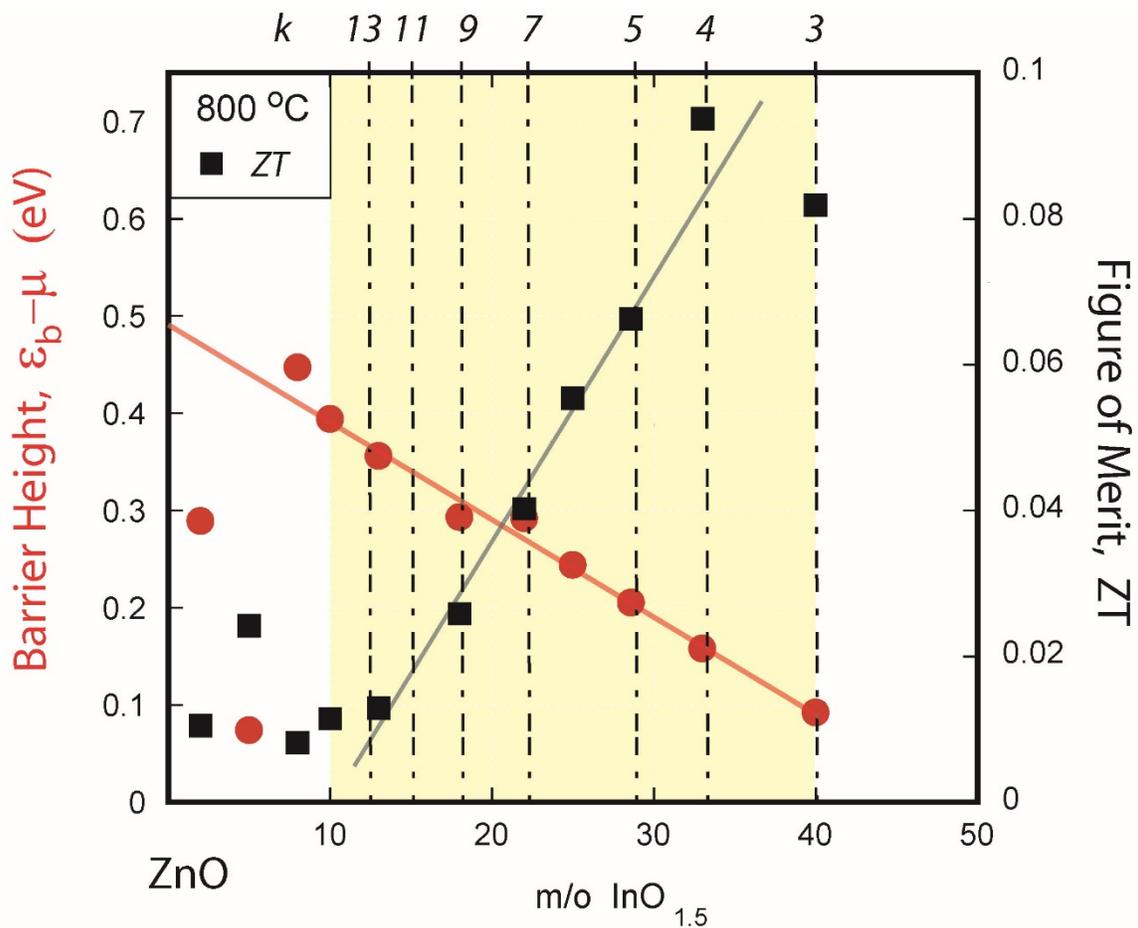

Figure 7: Electron potential barrier height of $In_2O_3(ZnO)_k$ superlattice interfaces relative to the Fermi level estimated according to equation 3 with input of Seebeck coefficients measured at 800 °C. Superimposed is the measured figure of merit ZT. The shaded region corresponds to the compositional range over which the natural superlattices form. The line through the barrier height data over the superlattice compositional range extrapolates to a barrier height of 0.49 eV.

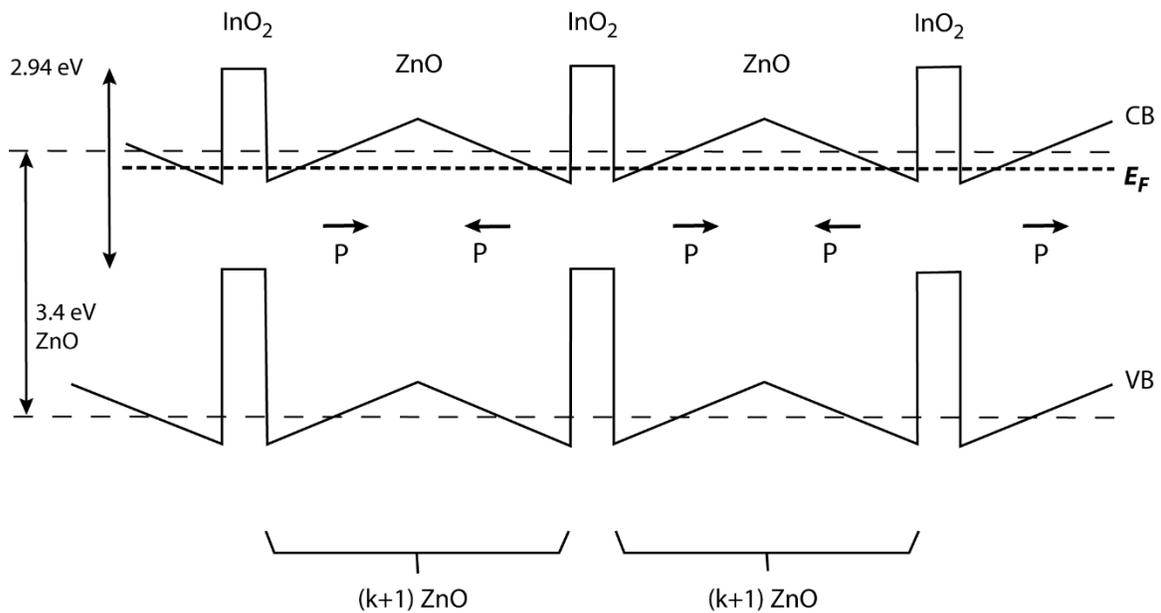

Figure 8. Proposed schematic band diagram for the $In_2O_3(ZnO)_k$ natural superlattices in the c-axis direction. The band offset between intrinsic $InO_2$ sheets and the ZnO provides a potential barrier to electron transport. The dashed lines represent the conduction and valence bands of the ZnO before polarization effects are taken into account. The crystallographic inversion at the $InO_2$ plane causes the spontaneous polarization to point away from the $InO_2$ sheets. Spontaneous polarization, P, produces internal electric fields shown schematically by the tilting of the bands whose magnitude increases as the size of the ZnO blocks decrease. Similarly, the crystallographic mirror plane midway between the $InO_2$ sheets produces a reversal in the electric field. Not shown is the local band bending in the ZnO adjacent to the $InO_2$ sheets where electrons will accumulate when the conduction band (CB) falls below the Fermi level. Not to scale.